\documentclass{article}
\usepackage{spconfa4,amsmath,graphicx,url,times}
\usepackage[nolist]{acronym}

\usepackage{color}
\usepackage{amsmath,amsfonts,amssymb,physics}
\usepackage{algorithm}
\usepackage{array}
\usepackage{textcomp}
\usepackage{stfloats}
\usepackage{float}
\usepackage{verbatim}
\usepackage{setspace}
\usepackage[noend]{algpseudocode}

\usepackage{caption}
\usepackage{subcaption}
\usepackage{xcolor}

\usepackage{comment}
\usepackage{cite}
\usepackage{soul}
\usepackage{epstopdf}
\usepackage{xfrac}
\usepackage{pifont}
\newcommand{\cmark}{\textcolor{green}{\ding{51}}}%
\newcommand{\xmark}{\textcolor{red}{\ding{55}}}%

\usepackage{csquotes}
\usepackage{booktabs}
\usepackage{multirow}
\usepackage{tabularx,ragged2e}
\newcolumntype{Y}{>{\centering\arraybackslash}X}
\usepackage{footmisc}

\usepackage{varwidth}
\usepackage{tikz}
\usepackage{pgf}
\usepackage{pgfplots}
\usepackage{pgfplotstable}
\usepackage{hyperref}
\usepackage{cleveref}

\usepackage{svg}
\usepackage{varwidth}
\usepackage{tikz}
\usepackage{pgf}
\usepackage{pgfplots}
\usepackage{pgfplotstable}
\usetikzlibrary{positioning,backgrounds,fit,calc,patterns,arrows.meta}
\pgfplotsset{compat=1.3}
\usetikzlibrary{shapes.geometric}
\usetikzlibrary{shapes.arrows}
\usetikzlibrary{arrows.meta}
\usepackage[skins]{tcolorbox}
\usepackage{adjustbox}

\usepackage{array}

\newcommand{\D}[0]{\mathrm{d}}
\newcommand{\state}[0]{\mathbf{x}_\tau}

\newcommand{\clean}[0]{\mathbf{x}_0}
\newcommand{\init}[0]{\mathbf{x}_{T_\mathrm{max}}}
\newcommand{\reverb}[0]{\mathbf{y}}
\newcommand{\sco}[0]{\nabla_{\state} \log p(\state)}

\newcommand{\rir}[0]{\mathbf{h}}
\newcommand{\postsco}[0]{\nabla_{\state} \log p(\state | \reverb, \rir)}
\newcommand{\likelihood}[0]{\nabla_{\state} \log p(\reverb | \state, \rir)}
\newcommand{\scomodel}[0]{\mathbf{s}_\theta(\state, \tau)}

\makeatletter
\newcommand{\subalign}[1]{%
  \vcenter{%
    \Let@ \restore@math@cr \default@tag
    \baselineskip\fontdimen10 \scriptfont\tw@
    \advance\baselineskip\fontdimen12 \scriptfont\tw@
    \lineskip\thr@@\fontdimen8 \scriptfont\thr@@
    \lineskiplimit\lineskip
    \ialign{\hfil$\m@th\scriptstyle##$&$\m@th\scriptstyle{}##$\hfil\crcr
      #1\crcr
    }%
  }%
}
\makeatother

\definecolor{statedps}{HTML}{FFC300}
\definecolor{dps}{HTML}{EC610A}
\definecolor{kodrasi}{HTML}{000000}

\title{BUDDy: Single-Channel Blind Unsupervised Dereverberation with Diffusion Models}

\name{Eloi Moliner$^{1\ast}$ \quad Jean-Marie Lemercier$^ 2$\sthanks{These authors contributed equally to this work.}\quad Simon Welker$^2$ \quad Timo Gerkmann$^2$ \quad Vesa Välimäki$^1$}
\address{$^1$ Acoustics Lab, Dept. Information and Communications Eng., Aalto University, Espoo, Finland \\ $^2$ Signal Processing (SP), Universität Hamburg, Germany}
\begin{acronym}
\acro{stft}[STFT]{short-time Fourier transform}
\acro{istft}[iSTFT]{inverse short-time Fourier transform}
\acro{dnn}[DNN]{deep neural network}
\acro{pesq}[PESQ]{Perceptual Evaluation of Speech Quality}
\acro{polqa}[POLQA]{perceptual objectve listening quality analysis}
\acro{wpe}[WPE]{weighted prediction error}
\acro{psd}[PSD]{power spectral density}
\acro{rir}[RIR]{room impulse response}
\acro{snr}[SNR]{signal-to-noise ratio}
\acro{lstm}[LSTM]{long short-term memory}
\acro{polqa}[POLQA]{Perceptual Objective Listening Quality Analysis}
\acro{sdr}[SDR]{signal-to-distortion ratio}
\acro{estoi}[ESTOI]{extended short-term objective intelligibility}
\acro{elr}[ELR]{early-to-late reverberation ratio}
\acro{tcn}[TCN]{temporal convolutional network}
\acro{rls}[RLS]{recursive least squares}
\acro{asr}[ASR]{automatic speech recognition}
\acro{ha}[HA]{hearing aid}
\acro{ci}[CI]{cochlear implant}
\acro{mac}[MAC]{multiply-and-accumulate}
\acro{vae}[VAE]{variational auto-encoder}
\acro{gan}[GAN]{generative adversarial network}
\acro{tf}[T-F]{time-frequency}
\acro{sde}[SDE]{stochastic differential equation}
\acro{ode}[ODE]{ordinary differential equation}
\acro{drr}[DRR]{direct to reverberant ratio}
\acro{lsd}[LSD]{log spectral distance}
\acro{sisdr}[SI-SDR]{scale-invariant signal to distortion ratio}
\acro{mos}[MOS]{mean opinion score}
\acro{map}[MAP]{maximum a posteriori}
\acro{sde}[SDE]{stochastic differential equation}
\acro{ode}[ODE]{ordinary differential equation}
\acro{dps}[DPS]{diffusion posterior sampling}
\end{acronym}

\begin{document}
\maketitle

\ninept
\maketitle

\begin{abstract}
In this paper, we present an unsupervised single-channel method for joint blind dereverberation and room impulse response estimation, based on posterior sampling with diffusion models. 
We parameterize the reverberation operator using a filter with exponential decay for each frequency subband, and iteratively estimate the corresponding parameters as the speech utterance gets refined along the reverse diffusion trajectory.
A measurement consistency criterion enforces the fidelity of the generated speech with the reverberant measurement, while an unconditional diffusion model implements a strong prior for clean speech generation.
Without any knowledge of the room impulse response nor any coupled reverberant-anechoic data, we can successfully perform dereverberation in various acoustic scenarios.
Our method significantly outperforms previous blind unsupervised baselines, and we demonstrate its increased
robustness to unseen acoustic conditions in comparison to blind supervised methods.
Audio samples and code are available online\footnote{\href{uhh.de/sp-inf-buddy}{uhh.de/sp-inf-buddy}.}.


\end{abstract}

\section{Introduction}

When acoustic waves propagate in enclosures and get reflected by walls, the sound received is perceived as reverberated, 
which can significantly degrade speech intelligibility and quality \cite{Naylor2011}. 
The goal of dereverberation is to recover the anechoic component from reverberant speech.
We focus here on the single-channel scenario, where measurements from only one microphone are available, which is significantly more challenging than multi-channel scenarios \cite{Miyoshi1988MINT}.

Traditional dereverberation algorithms assume some statistical properties, such as Gaussianity or sparsity, about the anechoic and reverberant signals. These properties are leveraged to perform dereverberation in the time, spectral or cepstral domain~\cite{gerkmann2018book_chapter}. 
These methods can tackle \textit{informed} scenarios, where the \ac{rir} is known \cite{Mourjopoulos1982Comparative, Kodrasi2014} as well as \textit{blind} scenarios where the \ac{rir} is unknown \cite{Nakatani2008a, yohena2024single}.
Informed dereverberation is easier than blind dereverberation,
but most scenarios in real-life applications are blind, as the \ac{rir} is either not measured beforehand, or becomes invalid even with the slightest deviations in receiver or emitter positions.

Data-driven approaches rely less on such assumptions but rather learn the signal properties and structures from data \cite{wang2018supervised}. 
Most of these methods are based on supervised learning using pairs of anechoic and reverberant speech.
Supervised predictive models have been widely used for blind dereverberation,
including \ac{tf} maskers~\cite{Williamson2017j}, time-domain methods~\cite{zhao2020}
and spectro-temporal mapping \cite{Han2017}.
Generative models represent another category of dereverberation algorithms aiming to learn the distribution of anechoic speech conditioned on reverberant input.
Some blind supervised methods using generative models such as diffusion models \cite{ho2020denoising, song2021sde}
have been recently proposed \cite{richter2023speech, Lemercier2022storm}. 
However, supervised approaches struggle with limited generalization to diverse acoustic conditions due to the scarcity and variability of available \ac{rir} data.
Unsupervised approaches offer the potential to circumvent such limitations as they do not require paired anechoic/reverberant data.
This paper builds upon prior work \cite{lemercier2023derevdps}, which proposed
 an unsupervised method for informed single-channel dereverberation based on diffusion posterior sampling.
The previous study showed the potential of leveraging diffusion models as a strong clean speech prior, which, when combined with a criterion to match the measurement, reached state-of-the-art dereverberation in an informed scenario \cite{lemercier2023derevdps}. This paper extends the method to 
blind dereverberation, where the 
unknown
\ac{rir}
is estimated along the anechoic speech.
We parameterize the \ac{rir} with a model-based subband filter, where each subband of the reverberation filter is modeled by an exponentially decaying signal. The resulting algorithm is an optimization scheme alternating between the diffusion process generating the anechoic speech, and the parameter search estimating the acoustic conditions.

Previous works in related domains explore various parameter estimation techniques for solving blind inverse problems with diffusion posterior sampling. 
For image deblurring, \cite{chung_parallel_2022} propose to use a parallel diffusion process to estimate the deblurring kernel, while \cite{laroche2023fast} adopts an expectation-maximization approach.
In the audio domain, \cite{moliner2024blind} address the problem of blind bandwidth extension by iteratively refining the parameters of the lowpass filter degradation.
Closely related is the work by Saito et al. \cite{saito2023unsupervised}, which perform unsupervised blind dereverberation using 
DDRM \cite{kawar2022ddrm} and the weighted-prediction error (WPE) algorithm as initialization \cite{Nakatani2008a}.

We name our method \textbf{BUDD}y for \textbf{B}lind \textbf{U}nsupervised \textbf{D}erever-beration with \textbf{D}iffusion Models. We show experimentally that BUDDy efficiently removes reverberation from speech utterances in many acoustic scenarios, thereby largely outperforming previous blind unsupervised techniques.
As supervision is not required during the training phase, we demonstrate that BUDDy does not lose performance when presented with unseen acoustic conditions,
as opposed to existing blind supervised dereverberation approaches.

\section{Background}
\subsection{Diffusion-Based Generative Models}
Diffusion-based generative models, or simply \textit{diffusion models} \cite{ho2020denoising, song2019generative}, emerged as a class of generative models that learn complex data distributions via iterative denoising. At training time, the target data distribution is transformed into a tractable Gaussian distribution by a \textit{forward process}, incrementally adding noise. During the inference, the \textit{reverse process} refines an initial noise sample into a data sample, by progressively removing noise.
The reverse diffusion process, which transports noise samples from a Gaussian prior to the data distribution $p_\text{data}$,
can be characterized by the following \emph{probability flow} \ac{ode}:
\begin{equation}\label{eq:ode}
    \D \state = [\mathbf{f}( \state,\tau)  - \tfrac{1}{2}g(\tau)^2  \sco ] \D \tau, 
\end{equation}
where $\tau$ indexes the diffusion steps flowing in reverse from $T_\mathrm{max}$ to $0$. The current diffusion state $\state$
starts from the initial condition $\init \sim \mathcal{N}(0,\sigma(T_\mathrm{max})^2 \mathbf{I})$ and ends at $\clean \sim p_\text{data}$. 
We adopt the variance exploding parameterization of Karras et al. \cite{karras2022elucidating}, where the \emph{drift} and \emph{diffusion} are defined as $f(\mathbf{\state}, \tau) = 0$ and $g(\tau) = \sqrt{2 \tau}$, respectively. 
Similarly, we adopt $\sigma(\tau)=\tau$ as the noise variance schedule, which defines the so-called \textit{transition kernel} i.e. the marginal densities:
$p_\tau(\state | \clean) = \mathcal{N}(\state; \clean, \sigma(\tau)^2 \mathbf{I})$.
The \emph{score function} $\sco$
is intractable at inference time as we do not have access to $\clean$. 
In practice, a \textit{score model} parameterized with a deep neural network $\scomodel$ is trained to estimate the score function
using a \emph{denoising score matching} objective \cite{vincent2011connection}.

\subsection{Diffusion Posterior Sampling for Dereverberation} \label{sec:dps}

Single-channel dereverberation can be considered as the inverse problem of retrieving the anechoic utterance $\clean \in \mathbb{R}^L$ from the reverberant measurement $\reverb \in \mathbb{R}^L$, which is often modelled by convolving the anechoic speech with an \ac{rir} $\rir \in \mathbb{R}^{L_\mathbf{h}}$, expressed as $\reverb = \rir \ast \clean$.
We aim to solve this inverse problem by sampling from the posterior distribution $p(\clean | \reverb, \rir)$ of anechoic speech given the measurement and the RIR. We adopt diffusion models for this posterior sampling task by replacing the score function $\sco$ in \eqref{eq:ode} by the \emph{posterior score} $\postsco$ 
\cite{song2021sde}. Applying Bayes' rule, the posterior score is obtained as\begin{equation} \label{eq:bayes}
    \postsco = \sco + \likelihood,
\end{equation}
where the first term, or \emph{prior score}, can be approximated with a trained score model  $\scomodel \approx \sco$. 
The likelihood $p(\reverb | \state, \rir)$ is generally intractable because we lack a signal model for $\reverb$ given the diffusion state $\state$. We will introduce in the next section a series of approximations to make its computation tractable.

\section{Methods}

\subsection{Likelihood Score Approximation}\label{sec:likelihood}

In order to obtain a tractable likelihood computation, we posit as in \cite{chung_diffusion_2022} that a one-step denoising estimate of $\clean$ at time $\tau$ can serve as a sufficient statistic for $\state$ in this context, i.e. that $p(\reverb | \state, \rir) \approx p(\reverb | \hat{\mathbf{x}}_0, \rir)$. 
Such estimate $\hat{\mathbf{x}}_0$ can be obtained
using the score model:
\begin{equation}
    \hat{\mathbf{x}}_0 
    \overset{\Delta}{=}
    \hat{\mathbf{x}}_0(\state, \tau) = 
    \state - \sigma(\tau)^2 \scomodel.
\end{equation}
Furthermore, we consider here that the convolution model remains valid when using this denoised estimate, and therefore that $p(\reverb | \hat{\mathbf{x}}_0, \rir) \approx p(\reverb | \hat{\mathbf{x}}_0  \ast \rir)$.
Finally, we model the estimation error as following a Gaussian distribution in the compressed STFT domain. 
\begin{equation} \label{eq:gaussian-likelihood}
    p(\reverb | \hat{\mathbf{x}}_0 \ast \rir) = \mathcal{N}(S_\text{comp}(\mathbf{y}); S_\text{comp}( \hat{\mathbf{x}}_0  \ast \rir) , \eta^2 \mathbf{I}),
\end{equation}
where $S_\text{comp}(\mathbf{y})=|\text{STFT}(\mathbf{y})|^{2/3} \exp{j \angle \text{STFT}(\mathbf{y})}$ is the compressed spectrogram. We apply this compression to 
account for the heavy-tailedness of speech distributions \cite{gerkmann2010}.
With this series of approximations, we obtain the following likelihood score:
\begin{equation}\label{eq:likelihood}
    \likelihood \approx - \zeta(\tau) \nabla_{\state} 
\mathcal{C}(\reverb, \rir \ast \hat{\mathbf{x}}_0),
\end{equation}
where the function $\mathcal{C}(\cdot, \cdot)$ is defined as:
\begin{equation}\label{eq:objective}
\mathcal{C}(\mathbf{y}, 
\hat{\mathbf{y}}
)
=
\frac{1}{M}\sum_{m=1}^M \sum_{k=1}^K 
\rVert
S_\text{comp}(\mathbf{y})_{m,k}-
S_\text{comp}(
\hat{\mathbf{y}}
)_{m,k}
\lVert_2^2.
\end{equation}
The weighting parameter $\zeta(\tau)$ controls the trade-off between adherence to the prior data distribution and fidelity to the observed data. According to our Gaussian assumption $\eqref{eq:gaussian-likelihood}$, its theoretical value
should depend on the unknown variance $\eta$ as $\zeta(\tau) = 1/2\eta^2$. In practice, we resort to the same parameterization
as in \cite{moliner_solving_2022, moliner2024blind}.

\subsection{Reverberation Operator}\label{sec:operator}

The employed reverberation operator relies on a subband filtering approximation 
\cite{avargel2007system}, which is applied within the Short-Time Fourier Transform (STFT) domain.
 Let $\mathbf{H} := \mathrm{STFT}(\mathbf{h}) \in \mathbb{C}^{N_\mathbf{h}\times K}$ represent the STFT of an RIR $\mathbf{h}$ with $N_\mathbf{h}$ time frames and $K$ frequency bins.
Similarly, let $\mathbf{X}\in \mathbb{C}^{M\times K}$, and $\mathbf{Y} \in \mathbb{C}^{M+N_\mathbf{h}-1\times K}$, denote the STFTs of anechoic $\mathbf{x}_0$ and reverberant $\mathbf{y}$ speech signals, repectively.
The subband convolution operation applies independent convolutions along the time dimension of each frequency band:
\begin{equation}\label{eq:subband}
\mathbf{Y}_{m,k}=\sum_{n=0}^{N_h}\mathbf{H}_{n,k} \mathbf{X}_{m-n,k}.
\end{equation}

In the blind scenario, we need to estimate $\mathbf{H}$, which is an arduous task without knowledge of the anechoic speech.
We constrain the space of possible solutions by 
designing a structured, differentiable \ac{rir} prior whose parameters $\psi$ can be estimated through gradient descent.
 We denote the complete forward reverberation operator, including forward and inverse STFT, as 
 $\mathcal{A}_\psi (\cdot): \mathbb{R}^{L} \rightarrow \mathbb{R}^{L}$.

We denote as $\mathbf{A} \in \mathbb{R}^{N_\mathbf{h}\times K}$ and $\mathbf{\Phi} \in \mathbb{R}^{N_\mathbf{h}\times K}$ the RIR magnitudes and phases of $\mathbf{H}$, respectively.
We parameterize the magnitude matrix $\mathbf{A}$ as a multi-band exponential decay model defined in $B<K$ frequency bands.
Let $\mathbf{A}^\prime \in \mathbb{R}^{N_\mathbf{h} \times B}$ be the subsampled version of $\mathbf{A}$ in the $B$ selected frequency bands.
Each frequency band $b$ is characterized by its weight $w_b$ and exponential decay rate $\alpha_b$, such that the corresponding subband magnitude filter can be expressed as:
\begin{equation}
\mathbf{A}_{n,b}^\prime= w_{b} e^{-\alpha_b n }.
\end{equation}
Once the weights and decay rates parameters are estimated, we reconstruct the magnitudes $\mathbf{A}$ by interpolating the subsampled $\mathbf{A}^\prime$ using
$\mathbf{A}=\exp(\mathrm{lerp}(\log(\mathbf{A}^\prime)))$, where $ \mathrm{lerp}$ represents linear interpolation of the frequencies.

Given the lack of structure of \ac{rir} phases, we perform independent optimization for each phase factor in $\mathbf{\Phi}$.
The resulting set of parameters to optimize is therefore $\psi = \{\mathbf{\Phi}, (w_b, \alpha_b)_{b=1,\dots, B} \}$.

After each optimization step, the estimated time-frequency RIR $\mathbf{H}$ is further processed through a projection step:
\begin{equation} \label{eq:projection}
    \overline{\mathbf{H}}=\mathrm{STFT}(\delta \oplus \mathcal{P}_\text{min}(\mathrm{iSTFT}(\mathbf{H}))).
\end{equation}
This operation primarily ensures STFT consistency \cite{gerkmann2015phase} of $\overline{\mathbf{H}}$.
We additionally include a projection $\mathcal{P}_\text{min}$ that ensures the time domain RIR has minimum phase lag to guarantee a stable inverse filter, using the Hilbert transform method \cite{oppenheimschafer}.
Finally, to make the direct-to-reverberation ratio only depend on the late reverberation and to enforce further constraints on $\psi$ for a more stable optimization, we take the direct path to be at the first sample and with amplitude one. This is achieved by replacing the first sample of the time-domain \ac{rir} with a unit impulse, as indicated by the operation $\delta \oplus ( \cdot )$.

\subsection{Blind Dereverberation Inference}\label{sec:blind_derev}
The inference process solves the following objective:
\begin{equation}
\label{eq:inverse_problem_optim}
  \hat{\mathbf{x}}_0, \hat{\psi}=  \underset{\mathbf{x}_0, \psi}{\mathrm{arg\,min}}\;  
    \mathcal{C}(\mathbf{y}, \mathcal{A}_\psi(\mathbf{x}_0))
    + \mathcal{R}(\psi),
     \quad \text{s.t.}\;\; \mathbf{x}_0 \sim p_{\text{data}}.
\end{equation}
This objective seeks to find the optimal speech $\hat{\mathbf{x}}_0$ and \ac{rir} parameters $\hat{\psi}$ that minimize the reconstruction error $\mathcal{C}(\mathbf{y}, \mathcal{A}_\psi(\mathbf{x}_0))$ while also incorporating a regularization term $\mathcal{R}(\psi)$. 
An essential aspect is the constraint $\mathbf{x}_0 \sim p_{\text{data}}$, which ensures that the estimated signal $\hat{\mathbf{x}}_0$ adheres to the distribution $p_{\text{data}}$ of anechoic speech samples. 
This constraint is implemented in a soft manner by leveraging a pre-trained score model $\mathbf{s}_\theta(\mathbf{x}_\tau, \tau)$ trained on anechoic speech.

\begin{figure}
    \resizebox{\columnwidth}{!}
    {
    \hspace{-2cm}
    \input{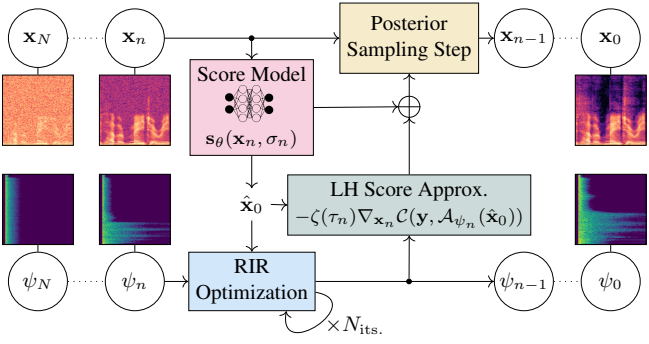}
    }
    \vspace{-1cm}
    \caption{\protect\textit{Blind unsupervised dereverberation alternating between RIR estimation and posterior sampling for speech reconstruction.}}
    \label{fig:diagram}
\vspace{-0.5cm}
\end{figure}

The inference algorithm is outlined in Algorithm \ref{alg:inference} and visualized in Fig.~\ref{fig:diagram}, using the discretization further described in Eq. \eqref{eq:discretization}.
The algorithm employs the likelihood score approximation from Sec.~\ref{sec:likelihood}, but replacing the convolution with the
the reverberation operator $\mathcal{A}_{\psi}(\cdot)$, 
while its parameters $\psi$ are optimized in parallel with the speech signal through gradient descent.

We introduce in \eqref{eq:inverse_problem_optim} a  \emph{noise regularization} term $\mathcal{R}(\psi)$:
\begin{equation}\label{eq:noise_reg}
\mathcal{R}(\psi)=
\frac{1}{N_\mathbf{h}}
\sum_{l=1}^{N_\mathbf{h}} \sum_{k=1}^K 
\rVert
S_\text{comp}(
\hat{\mathbf{h}}_\psi)_{l,k}
-
S_\text{comp}(\hat{\mathbf{h}}_{\psi^\prime}+\sigma^\prime\mathbf{v})_{l,k}
\lVert_2^2,
\end{equation}
where $\hat{\mathbf{h}}_\psi=\mathcal{A}_\psi(\delta)$ 
represents the estimated RIR in the waveform domain, $\mathbf{v} \sim \mathcal{N}(\mathbf{0}, \mathbf{I})$ is a vector of white Gaussian noise,
and
$\hat{\mathbf{h}}_{\psi^\prime}$ is a copy of the current estimate of $\hat{\mathbf{h}}_\psi$,
such that the $\mathrm{arg\,min}$ in \eqref{eq:inverse_problem_optim} does not apply to it.
In code, this is analogous to detaching the gradients of $\hat{\mathbf{h}}_\psi$ using a $\mathrm{stop\;grad}$ operator.
We adopt an annealed schedule for the noise level $\sigma^\prime(\tau)$, resembling the score model schedule $\sigma(\tau)$ but with different hyper-parameters.
This regularization term injects noise in the \ac{rir} parameter gradients, with decreasing noise power, which enables a wider and smoother exploration while allowing for convergence toward the end of the optimization.

\begin{table*}[h]
    \centering
    \caption{\centering\textit{Dereverberation results obtained on VCTK-based reverberant datasets. Values indicate mean and standard deviation. We indicate for each method in the table if is blind (i.e. have no knowledge of the RIR) and/or unsupervised. Boldface numbers indicate best performance for supervised and unsupervised methods separately. For all metrics, higher is better.}}
    \scalebox{0.94}{
    \begin{tabular}{l|c|c|ccc|ccc}
    
 \multicolumn{3}{c}{} & \multicolumn{3}{c}{Matched} &  \multicolumn{3}{c}{Mismatched} \\
\cmidrule(lr){4-6} \cmidrule(lr){7-9}
Method & Blind & Unsup. & DNS-MOS & PESQ & ESTOI & DNS-MOS & PESQ & ESTOI \\

\midrule
\midrule
Reverberant & - & - 
& 3.14 $\pm$ 0.52 & 1.61 $\pm$ 0.37 & 0.50 $\pm$ 0.14 
& 3.05 $\pm$ 0.47 & 1.57 $\pm$ 0.29 & 0.47 $\pm$ 0.11 \\
\midrule

RIF+Post \cite{Kodrasi2014} & \xmark & \cmark 
& 3.41 $\pm$ 0.47 & 2.66 $\pm$ 0.40 & 0.76 $\pm$ 0.09
& 3.55 $\pm$ 0.45 & 2.86 $\pm$ 0.31 & 0.78 $\pm$ 0.09 \\
 
InfDerevDPS \cite{lemercier2023derevdps} & \xmark & \cmark 
& 3.91 $\pm$ 0.35 & 3.77 $\pm$ 0.41 & 0.83 $\pm$ 0.09
 & 3.92 $\pm$ 0.32 & 3.69 $\pm$ 0.31 & 0.84 $\pm$ 0.08 \\
\midrule \midrule

NCSN++M \cite{Lemercier2022analysing} & \cmark & \xmark
& 3.75 $\pm$ 0.38 & 2.85 $\pm$ 0.55 & 0.80 $\pm$ 0.10 
& 3.61 $\pm$ 0.39 & 2.08 $\pm$ 0.47 & 0.64 $\pm$ 0.09 \\

SGMSE+M \cite{richter2023speech, Lemercier2022analysing} & \cmark & \xmark
& 3.88 $\pm$ 0.32 & 2.99 $\pm$ 0.48 & 0.78 $\pm$ 0.09
 & 3.74 $\pm$ 0.34 & 2.48 $\pm$ 0.47 & \textbf{0.69} $\pm$ \textbf{0.09} \\
 
StoRM \cite{Lemercier2022storm} & \cmark & \xmark
& \textbf{3.90 $\pm$ 0.33} & \textbf{3.33 $\pm$ 0.48} & \textbf{0.82 $\pm$ 0.10}
& \textbf{3.83} $\pm$ \textbf{0.32} & \textbf{2.51} $\pm$ \textbf{0.53} & 0.67 $\pm$ 0.09 \\ 
\midrule
\midrule

Yohena and Yatabe \cite{yohena2024single} & \cmark & \cmark 
& 2.99 $\pm$ 0.56 & 1.80 $\pm$ 0.33 & 0.55 $\pm$ 0.12 
& 2.94 $\pm$ 0.44 & 1.71 $\pm$ 0.29 & 0.51 $\pm$ 0.10 \\

WPE \cite{Nakatani2008b} & \cmark & \cmark 
& 3.24 $\pm$ 0.54 & 1.81 $\pm$ 0.42 & 0.57 $\pm$ 0.14
& 3.10 $\pm$ 0.48 & 1.74 $\pm$ 0.37 & 0.54 $\pm$ 0.12 \\
 
Saito et al. \cite{saito2023unsupervised} & \cmark & \cmark 
& 3.22 $\pm$ 0.56 & 1.68 $\pm$ 0.40 & 0.51 $\pm$ 0.13 
& 3.12 $\pm$ 0.52 & 1.70 $\pm$ 0.33 & 0.52 $\pm$ 0.10 \\

GibbsDDRM \cite{murata_gibbsddrm_2023} & \cmark & \cmark
& 3.33 $\pm$ 0.53 & 1.70 $\pm$ 0.37 & 0.51 $\pm$ 0.13
& 3.30 $\pm$ 0.52 & 1.75 $\pm$ 0.36 & 0.52 $\pm$ 0.11 \\

BUDDy (proposed) & \cmark & \cmark 
& \textbf{3.76} $\pm$ \textbf{0.41} & \textbf{2.30} $\pm$ \textbf{0.53} & \textbf{0.66} $\pm$ \textbf{0.12}
& \textbf{3.74} $\pm$ \textbf{0.38} & \textbf{2.24} $\pm$ \textbf{0.54} & \textbf{0.65} $\pm$ \textbf{0.12} \\

\midrule
    \bottomrule
    \end{tabular}
    }
    \label{tab:results}
\end{table*}

\definecolor{cb1}{HTML}{D81B60}
\definecolor{cb2}{HTML}{1E88E5}
\definecolor{cb3}{HTML}{D29E02}
\definecolor{cb4}{HTML}{004D40}

\begin{algorithm}[t]
\caption{Inference algorithm}
\label{alg:inference}
\begin{algorithmic}
\Require reverberant speech $\mathbf{y}$
\State $\mathbf{x}_\mathrm{init} \leftarrow \mathrm{WPE}(\mathbf{y})$  
\State Sample $\mathbf{x}_{N}\sim \mathcal{N}({\mathbf{x}_\mathrm{init}},\sigma_N^2\mathbf{I})$ \Comment{Warm initialization}
\State Initialize $\psi_{N}$ \Comment{Initialize the RIR parameters}
\For{$n \leftarrow N, \dots, 1$} \Comment{Discrete step backwards}
\State $\mathbf{s}_n \leftarrow s_\theta(\mathbf{x}_n, \tau_n) $ \Comment{\textcolor{cb1}{Evaluate score model}}
\State $\hat{\mathbf{x}}_0 \leftarrow  \mathbf{x}_n - \sigma_n^2 \mathbf{s}_n$ \Comment{Get one-step denoising estimate}
\State $\hat{\mathbf{x}}_0 \leftarrow \mathrm{Rescale}(\hat{\mathbf{x}}_0) $

\State $\psi_{n-1}^0 \leftarrow \psi_{n}$  \Comment{Use the RIR parameters from last step}
\For{$j \leftarrow 0, \dots, N_\text{its.}$} \Comment{\textcolor{cb2}{RIR optimization}}
    \State $\mathcal{J}_\text{RIR}(\psi_{n-1}^j) \leftarrow \mathcal{C}(\mathbf{y},\mathcal{A}_{\psi_{n-1}^j}(\hat{\mathbf{x}}_0)) + \mathcal{R}(\psi_{n-1}^j)$
\State $\psi_{n-1}^{j+1} \leftarrow \psi_{n-1}^{j} - \mathrm{Adam}(\mathcal{J}_\text{RIR}(\psi_{n-1}^{j}))$  \Comment{Optim. step}
 \State $\psi_{n-1}^{j+1} \leftarrow \mathrm{project}(\psi_{n-1}^{j+1})$ \Comment{Projection step}
\EndFor
\State $\psi_{n-1} \leftarrow \psi_{n-1}^M$
\State $ \mathbf{g}_n \leftarrow \zeta(\tau_n) \nabla_{\mathbf{x}_n} 
    \mathcal{C}(\reverb,
    \mathcal{A}_{\psi_{n-1}}(\hat{\mathbf{x}}_0
    ))$ \Comment{\textcolor{cb4}{LH score approx.}}
\State $\mathbf{x}_{{n-1}} \leftarrow \mathbf{x}_{n} - \sigma_n (\sigma_{n-1}-\sigma_n) (\mathbf{s}_n +\mathbf{g}_n)$ \Comment{\textcolor{cb3}{Update step}}
\EndFor
\State \Return $\mathbf{x}_0$  \Comment{Reconstructed audio signal}
\end{algorithmic}
\end{algorithm}

\section{Experimental Setup}

\subsection{Data}
\label{sec:exp:data}

We use VCTK \cite{valentini2016reverberant} as clean speech, selecting 103 speakers for training, 2 for validation and 2 for testing. 
We curate recorded \acp{rir} from various public datasets (please visit our code repository
for details).
In total we obtain approximately 10,000 RIRs, and split them between training, validation, and testing using ratios 0.9, 0.05, and 0.05, respectively. The training and validation sets are only used to train the baselines which require coupled reverberant/anechoic data.
All data is resampled at 16 kHz.

\subsection{Baselines}
\label{sec:exp:baselines}

We compare our method BUDDy to several blind supervised baselines such as NCSN++M \cite{Lemercier2022analysing} and diffusion-based SGMSE+ \cite{richter2023speech} and StoRM \cite{Lemercier2022storm}. We also include blind unsupervised approaches leveraging traditional methods such as WPE \cite{Nakatani2008a} and Yohena et al. \cite{yohena2024single}, as well as diffusion models Saito et al. \cite{saito2023unsupervised} and GibbsDDRM \cite{murata_gibbsddrm_2023} with code provided by the authors. 
For WPE, we take 5 iterations, a filter length of 50 STFT frames (400\,ms) and a delay of 2 STFT frames (16\,ms).

\subsection{Hyperparameters and Training Configuration}
\label{sec:exp:hyperparameters}

\textit{Data representation}:
\label{sec:hyperparameters-data}
We train the score model $\mathbf{s}_\theta$ using only the anechoic data from VCTK. 
For training, 4-s segments are randomly extracted from the utterances. 
Using publicly available code, the blind supervised models NCSN++M \cite{Lemercier2022analysing}, SGMSE+ \cite{richter2023speech} and StoRM \cite{Lemercier2022storm} are trained using coupled reverberant/anechoic speech, where the reverberant speech is obtained by convolving the anechoic speech from VCTK with the normalized RIRs.

\newcommand{\vv}[0]{0.35em}

\vspace{\vv}
\noindent\textit{Reverberation operator}:
\label{sec:hyperparameters-operator}
For all methods, STFTs are computed using a Hann window of 32 ms and a hop size of 8 ms. For subband filtering, we further employ 50\% zero-padding to avoid aliasing artifacts. Given our sampling rate of $f_\text{s}=16$ kHz, this results in $K=513$ frequency bins. We set the number of STFT frames of our operator to $N_\mathbf{h} = 100$ (800 ms).
We subsample the frequency scale in $B=26$ bands, with a $125$-Hz spacing between $0$ and $1$ kHz, a $250$-Hz spacing between $1$ and $3$ kHz, and a $500$-Hz spacing between $3$ and $8$ kHz.
We optimize the \ac{rir} parameters $\psi$ with Adam, where the learning rate is set to 0.1, the momentum parameters to $\beta_1=0.9$, and $\beta_2=0.99$, and $N_\text{its.}=10$ optimization iterations per diffusion step.
We constrain the weights $w_b$ between 0 and 40\;dB, and the decays $\alpha_b$ between 0.5 and 28. This prevents the optimization from approaching degenerate solutions at early sampling stages.
Furthermore, we rescale the denoised estimate $\hat{\mathbf{x}}_0$ at each step to match the empirical dataset standard deviation $\sigma_\mathrm{data} = 5 \cdot 10^{-2}$, so as
to enforce a constraint on the absolute magnitudes of $\hat{\mathbf{h}}_\psi$ and $\hat{\mathbf{x}}_0$.

\vspace{\vv}
\noindent\textit{Forward and reverse diffusion} 
\label{sec:hyperparameters-diffusion}
We set the extremal diffusion times to $T_\mathrm{max}=0.5$ and $T_\mathrm{min}=10^{-4}$. 
For reverse diffusion, we follow Karras et al. \cite{karras2022elucidating} and employ a discretization of the diffusion time axis using $N=200$ steps according to:
\begin{equation}\label{eq:discretization}
    \forall n < N,  \, \tau_n = \sigma_n = \left(T_\mathrm{max}^{1/\rho} + \frac{n}{N-1} ( T_\mathrm{min}^{n/\rho} - T_\mathrm{max}^{1/\rho} )\right)^\rho,
\end{equation}
with warping $\rho=10$. We use the second-order Euler-Heun stochastic sampler in \cite{karras2022elucidating} with $S_\mathrm{churn}=50$ and $\zeta^\prime=0.5$ (prior scaling, see \cite{moliner_solving_2022}), and the initial point $\mathbf{x}_\mathrm{init}$ is taken to be the output of WPE \cite{Nakatani2008a} (with same parameters as the WPE baseline) plus Gaussian noise with standard deviation $\sigma = T_\mathrm{max}$. 
The annealing schedule $\sigma^\prime(\tau)$ in the noise regularization term in \eqref{eq:noise_reg} is the same as the diffusion noise schedule $\sigma(\tau)$ but we bound it between extremal values $\sigma_\text{min}^\prime=5\times 10^{-4}$ and $\sigma_\text{max}^\prime=10^{-2}$.

\vspace{\vv}
\noindent\textit{Network architecture}: 
\label{sec:hyperparameters-network}
To remain consistent with \cite{lemercier2023derevdps}, the unconditional score network architecture is NCSN++M\cite{Lemercier2022analysing,Lemercier2022storm}, a lighter variant of the NCSN++ \cite{song2021sde} with 27.8M parameters instead of 
65M.

\vspace{\vv}
\noindent\textit{Training configuration}:\label{sec:hyperparameters-training}
We adopt Adam as the optimizer to train the unconditional score model, 
with a learning rate of $10^{-4}$ and an effective batch size of 16 for 190k steps. 
We track an exponential moving average of the DNN weights with a decay of 0.999.

\vspace{\vv}
\noindent\textit{Evaluation metrics}:
\label{sec:hyperparameters-eval}
We assess the quality and intelligibility of speech using the intrusive \ac{pesq} \cite{Rix2001PESQ} and \ac{estoi} \cite{Jensen2016ESTOI}. We also employ the non-intrusive DNS-MOS \cite{reddy2021dnsmos}, as a DNN-based \ac{mos} approximation.

\section{Results and Discussion}

Table \ref{tab:results} shows the dereverberation results for all baselines and indicates whether each approach is blind and/or unsupervised. We included the results for RIF+Post \cite{Kodrasi2014} and InfDerevDPS \cite{lemercier2023derevdps} in the informed scenario to show the upper bound of dereveberation quality one can achieve with perfect knowledge of the room acoustics.
We use the same score model $\mathbf{s}_\theta$ and cost function $\mathcal{C}(\cdot, \cdot)$ for InfDerevDPS \cite{lemercier2023derevdps} as for BUDDy.
Blind supervised approaches NCSN++M, SGMSE+M, and StoRM largely profit from the supervision during training, and boast a better performance compared to the unsupervised methods. 
However, in the mismatched setting, their performance dwindles because of their limited generalizability.
In contrast, the proposed method BUDDy benefits from unsupervised training, and therefore, modifying the acoustic conditions does not impact performance at all: typically NCSN++M loses 0.78 PESQ by switching from the matched case to the mismatched case, where BUDDy loses 0.06. 
Our method then outperforms NCSN++M and comes within reach of other supervised approaches, although the generative nature of SGMSE+ and StoRM allow them to retain a relatively high generalization ability.
We also observe that the traditional blind unsupervised methods such as WPE \cite{Nakatani2008a} and Yohena and Yatabe \cite{yohena2024single} can only perform limited dereverberation, as they do not benefit from the strong anechoic speech prior that learning-based methods parameterized with deep neural networks offer.
Finally, we note that BUDDy performs significantly better on all metrics than the diffusion-based blind unsupervised baselines Saito et al. \cite{saito2023unsupervised} and GibbsDDRM \cite{murata_gibbsddrm_2023}, as these perform mild dereverberation in the presented acoustic conditions, where the input direct-to-reverberant ratio is significanty lower than in the authors' setup.

\section{Conclusions}
This paper presents BUDDy, the first unsupervised method simultaneously performing blind dereverberation and \ac{rir} estimation using diffusion posterior sampling. 
BUDDy significantly outperforms traditional and diffusion-based unsupervised blind approaches. 
Unlike blind supervised methods, which often struggle with generalization to unseen acoustic conditions, our unsupervised approach overcomes this limitation due to its ability to adapt the reverberation operator to a broad range of room impulse responses.
While blind supervised methods outperform our approach when the tested conditions match those at training time, our method is on par or even outperforms some supervised baselines in a mismatched setting.

\bibliographystyle{IEEEbib}
\bibliography{refs23abbreviated}

\appendix

\end{document}